# Large uncertainties in the thermodynamics of phosphorus (III) oxide ($P_4O_6$) have significant implications for phosphorus species in planetary atmospheres.

*William Bains[1,2,3] *, Matthew A. Pasek[4], Sukrit Ranjan[5], Janusz J. Petkowski[1,6], Arthur Omran[4], Sara Seager[1,7,8]*

[1] Department of Earth, Atmospheric and Planetary Sciences, Massachusetts Institute of Technology, 77 Massachusetts Avenue, Cambridge, MA 02139, USA

[2] School of Physics and Astronomy, Cardiff University, 4 The Parade, Cardiff CF24 3AA, UK

[3] Rufus Scientific, Melbourn, Royston, Herts, SG8 6ED, UK

[4] Department of Geosciences, University of South Florida, Tampa, FL 33620, USA

[5] Lunar & Planetary Laboratory/Department of Planetary Sciences, University of Arizona, Tucson, AZ 85721-0092, US

[6] JJ Scientific, 02-792 Warsaw, Poland

[7] Department of Physics, Massachusetts Institute of Technology, 77 Massachusetts Avenue, Cambridge, MA 02139, USA

[8] Department of Aeronautics and Astronautics, Massachusetts Institute of Technology, 77 Massachusetts Avenue, Cambridge, MA 02139, USA

* Corresponding author: bains@mit.edu

ABSTRACT Phosphorus (III) oxide ($P_4O_6$) has been suggested to be a major component of the gas phase phosphorus chemistry in the atmospheres of gas giant planets and of Venus. However, $P_4O_6$'s proposed role is based on thermodynamic modelling, itself based on values for the free energy of formation of $P_4O_6$ estimated from limited experimental data. Values of the standard Gibbs free energy of formation ($\Delta G°_{(g)}$) of $P_4O_6$ in the literature differ by up to ~656 kJ/mol, a huge range. Depending on which value of is assumed, $P_4O_6$ may either be the majority phosphorus species present or be completely absent from modelled atmospheres. Here we critically review the literature thermodynamic values and compare their predictions to observed constraints on $P_4O_6$ geochemistry. We conclude that the widely used values from the NIST/JANAF database are almost certainly too low (predicting $P_4O_6$ is more stable than is plausible). We show that, regardless of the value of $\Delta G°_{(g)}$ for $P_4O_6$ assumed, the formation of phosphine from $P_4O_6$ in the Venusian atmosphere is thermodynamically unfavorable. We conclude that there is a need for more robust data on both the thermodynamics of phosphorus chemistry for astronomical and geological modelling in general, and for understanding the atmosphere of Venus and the gas giant planets in particular.

# TEXT

## 1. Background

"My advice to you, sir? Always verify your references!" [1]

This study was initiated to understand the one of the major differences between Bains et al.'s [2] and Omran et al.'s [3] conclusions about the possible source of phosphine ($PH_3$) in the atmosphere of Venus. Bains et al. [2] predicted that phosphine ($PH_3$) would not spontaneously form in Venus' atmosphere by disproportionation of $P_4O_6$, Omran et al. [3] predicted that it would. Phosphine was proposed to be formed when phosphorus (III) oxide ($P_4O_6$) reacted with water to transiently form phosphorous acid ($H_3PO_3$) which then undergoes disproportionation to phosphine and phosphoric acid ($H_3PO_4$), as phosphorous acid ($H_3PO_3$) itself is not stable in gas phase:

$$P_4O_6 + 6\ H_2O \rightarrow [4\ H_3PO_3] \rightarrow PH_3 + 3\ H_3PO_4 \qquad (1)$$

Bains et al. [2] found that this reaction's thermodynamics predicted that $P_4O_6$ would not spontaneously generate 1 ppb $PH_3$ in Venus' atmosphere, and $P_4O_6$ was argued to be the primary gaseous P compound in the lower atmosphere (0-20 km) of Venus. Omran et al. [3] found that the reaction was highly exergonic and would generate $PH_3$ if $P_4O_6$ was present, but did not determine the equilibrium P chemistry of Venus's atmosphere to show whether $P_4O_6$ was present at low altitudes. We identified thermodynamic values, and particularly the free energy of formation of $P_4O_6$, as key differences between the two studies. The Gibbs free energy of a reaction (ΔG) is directly related to the equilibrium of that reaction, such that

$$\Delta G = -R \cdot T \cdot \ln(K)$$

Where ΔG is the free energy change of a reaction, R is the gas constant, T is the absolute temperature, and K is the equilibrium constant, which is defined as

$$K = \frac{\prod_{i=1-n}\{P_i\}^{\alpha_i}}{\prod_{j=1-m}\{R_j\}^{\alpha_j}}$$

Where {P} are the activities of the products, {R} are the activities of the reactants, $a_i$ and $a_j$ are the number of moles of each product and reactant respectively in the reaction. Activity is approximately the same as concentration in dilute gas phase far from the substance's critical point. Thus, if the free energy is negative, the products of a reaction will be favored over reactants at equilibrium. The free energy of formation of a substance (ΔG°) is the free energy of a notional reaction forming that substance from its elements, at 101 kPa and 298 K. Bains et al.[2] used data from the NIST/JANAF tables [4, 5]. Omran et al.[3] used data from the HSC database, provided by Mogroup[6]. The differences between these 'standard' values are summarized in Table 1.

|  | ΔG° (kJ/mol) | |
| --- | --- | --- |
| Substance | NIST/JANAF [4, 5] (assumed by Bains et al. [2]) | HSC [6] (assumed by Omran et al. [3]) |
| $PH_3(g)$ | 30.893 | 13.45 |
| $P_4O_6(g)$ | -2085 | -1480 |
| $H_3PO_4(l)$ | -1123.6 | -1149 |
| $H_2O(g)$ | -228.6 | -228.6 |

**Table 1.** There are substantial differences in the ΔG° of $P_4O_6$ between different sources, as well as differences in the ΔG°. of other species.

While there are minor differences in two of the values, there is a very large difference in the $\Delta G°_{(g)}$ of $P_4O_6$ (shaded in Table 1). The same difference is probably the cause of the dispute between Borunov et al. [7], who claim that $PH_3$ is in equilibrium in Jupiter's 1-bar level atmosphere using the higher $\Delta G°_{(g)}$ for $P_4O_6$ from [8], and Visscher et al. [9] who claim that $PH_3$ is not an equilibrium component of the 1-bar atmosphere using the thermodynamic values from NIST/JANAF [4,5]. Wang et al [4,5,10] also modelled the atmospheres of Jupiter and Saturn and found that $H_3PO_4$ was the dominant species up to a temperature of 700 K, unlike the literature consensus [9,11] which predicts $P_4O_6$ is dominant above ~500 K; Wang et al [10] identify the difference between the $\Delta H°$ derived from NIST/JANAF [4] and from Hartley and McCoubrey [8] as the source of the different predictions.

Here we seek to identify where the difference in $\Delta G°_{(g)}$ for $P_4O_6$ comes from, and whether the difference in values affects the possibility of a source for phosphine on Venus derived from abiotic mechanisms dominated by thermodynamics. We conclude by discussing the broader impact of the differences in the thermodynamic values for $P_4O_6$ on planetary science, and what is the range in which the most accurate value is likely to lie.

## 2. Sources of data on $P_4O_6$ thermodynamics

### 2.1 Published experimental values

There is little published experimental data on $P_4O_6$. A search of Google Scholar [12] for "phosphorus trioxide" or "phosphorus (III) oxide" gave only ~360 hits covering papers from the last 120 years, and only two relating to measurements of its thermodynamics. W.B. asked HSC for the source of their values. HSC stated that reference [13] gave a value for $\Delta G°_{(l)}$ of $P_4O_6$ as -2099.630 kJ/mol, and reference [14] gave a value for $\Delta G°_{(g)}$ of -1479.718 kJ/mol. If both these values are accurate, then the Gibbs free energy of evaporation of $P_4O_6$ would be ~620 kJ/mol. This is an extraordinarily high value, comparable to the energy needed to evaporate diamond. One of these values must therefore be in error.

The NIST/JANAF tables take their value for the solid phase $\Delta G°_{(s)}$ of $P_4O_6$ from [15]. That paper states that the measured enthalpy of formation ($\Delta H°$) of solid phase "phosphorus trioxide" at 298 K is -270 kCal/mol of $P_2O_3$, which is equivalent to -2270.87 kJ/mol $P_4O_6$. No value for gas phase is given. No value for the solid phase entropy is given. Hartley and McCoubrey [8] give a value of $\Delta H°_{(s)}$ -1671.1 kJ/mol; however Chase [4], which is the basis of the NIST/JANAF tables, did not choose this value. Foote [16] gives $\Delta H°$, as -1622.1 kJ/mol, but gives no source for this value, or what phase it applies to. Heinz et al [17] give a value for $\Delta H°_{(g)}$ of $P_4O_6$ as -1600 kJ/mol. A review of the Soviet-era Russian literature reports the enthalpy of formation for $P_4O_6$ to be -2179 kJ/mol for solid phase, -2108 kJ/mol for gas phase [18], but without experimental details.

The original IVANTHERMO database [14] is not readily available. OSTI [19] mention the database and say that it is more up to date than NIST/JANAF, but that it is incomplete.

## 2.2 Estimating gas phase enthalpy from solid phase enthalpy and entropies

At the boiling or sublimation temperature of $P_4O_6$, we would expect the $\Delta G$ of the interconversion of gas and condensed phases to be 0. Therefore,

$P_4O_{6(s)} \rightarrow P_4O_{6(g)}$

$\Delta G°_{(g)} - \Delta G°_{(s)} = 0$

$\Delta G° = \Delta H° - T\Delta S°$

$$\Delta H°_{(s)} + T \cdot \Delta S°_{(s)} = \Delta H°_{(s)} + \Delta H_e + T \cdot \Delta S°_{(g)}$$

where $\Delta H_e$ is the heat of evaporation, T is the absolute temperature, $\Delta G°$ is the standard free energy of formation, and $\Delta S°$ is the difference between the entropy of the substance at that temperature and the entropy of the elements of which it is composed at the same temperature. Hence

$$\Delta H_e = T(\Delta S°_{(g)} - \Delta S°_{(s)})$$

i.e., the temperature T where the solid sublimes or a liquid boils is that at which the entropy change between condensed phase and gas balances the heat of evaporation $\Delta H_e$. A larger $\Delta H_e$ implies that a higher temperature is needed to achieve the same entropy change between condensed and gas phase.

From this logic, we can calculate the gas phase free energy of formation of $P_4O_6$ from the measured solid phase values and the entropies of solid and gas phases derived from NIST/JANAF [5]. For this we have to assume a constant $\Delta H$ with temperature. This is not accurate, but the $\delta\Delta H/\delta T$ is only available for gas phase $P_4O_6$ from NIST/JANAF, which we cannot use as the NIST/JANAF data for enthalpy are in dispute, and $\delta\Delta H/\delta T$ for the solid phase is not available at all. We have no data for the specific heat capacity of solid $P_4O_6$. Therefore, this is an estimate, not an accurate calculation.

The $\Delta S°_{(g)}$ for $P_4O_6$ from NIST/JANAF [5] at 298 K is -345.641 J/mol/K. We have also carried out extensive modelling of the entropy of molecules using semi-empirical QM calculations, which predict gas phase entropy across a diverse set of molecules with an RMS error of 12.85 J/mol/K [20]. We used this method to predict the entropy of $P_4O_{6(g)}$ and $P_4O_{10(g)}$, as summarized in Table 2.

| Substance | Entropy of formation (J/mol/K) | | |
|---|---|---|---|
| | 298 K | 444 K (171°C) | 633 K (360 °C) |
| $P_4O_{10\,(s)}$ | -961.262 | -972.961 | -955.75 |
| $P_4O_{10\,(g)}$ | -786.079 | -808.865 | -810.421 |
| $P_4O_{6\,(g)}$ | -434.108 | -453.778 | |

**Table 2.** Calculated entropy of formation of phosphorus (III) oxide and phosphorus (V) oxide, in gas phase from QM estimates as described in [20] and in solid phase from NIST/JANAF.

The entropy of solid $P_4O_6$ is not known. As a rough approximation, entropy scales with the number of atoms in a molecule (at least for molecules of similar topology), so in principle we can scale the solid phase entropy value of $P_4O_{10}$ by 10/14 to estimate the solid phase entropy of $P_4O_6$. Assuming the $\Delta S°_{(s)}$ of $P_4O_6$ = -694.972 J/mol/K, then at sublimation temperature (444 K)

$\Delta H_e = 444 \cdot (-453.778 + 694.972)$ J/mol = 107.09 kJ/mol

And hence

$\Delta G°_{(g)} = \Delta H°_{(s)} + \Delta H_e - T\Delta S°_{(g)} = -2028.55$ kJ/mol

assuming the enthalpy of formation of the solid from [15] used by NIST/JANAF, and -1428.78 assuming the enthalpy of formation of the solid from ref [8].

## 2.3 Prediction from QM calculations

We have also used our semi-empirical modelling approach to predict $\Delta G°_{(g)}$ for $P_4O_6$ and $P_4O_{10}$ (these two molecules are not included in the "All Small Molecules" (ASM) set [21] as they have more than 6 non-H atoms, but the same methods were used as for the ASM thermodynamic calculations [20]). This method is different from and independent of NIST/JANAF, and was validated on a much wider dataset than is available in NIST/JANAF (see [20] for details). Notably, $P_4O_6$ was not in the training set for model development. The predicted ΔH and ΔS values, and the ΔG values that can be calculated from them at 298K, are given in table 3.

|  | $\Delta H^0_{(g)}$ (kJ/mol) | $\Delta S^o_{(g)}$ (J/mol/K) | $\Delta G^o_{(g)}$ (kJ/mol) |
|---|---|---|---|
| $P_4O_6$ | -1985 | -420.15 | -1859.8 |
| $P_4O_{10}$ | -2907 | -794.54 | -2670.22 |

**Table 3.** Enthalpy, Entropy and Gibbs free energy of formation of gas phase phosphorus (III) oxide and phosphorus (V) oxide, as calculated using semi-empirical quantum mechanical methods as described in [20].

Morgon [22] has used ab initio quantum mechanical methods, using a modified version of the correlation consistent composite approach, to derive a ΔH° for $P_4O_6$ of −1706.3 kJ/mol (presumed to be in gas phase).

## 2.4 Calculation from energy of atomization

We can calculate the enthalpy of formation of $P_4O_6$ in the gas phase from the energy of atomization, which can be derived directly by mass spectrometry. This has been done by [23], and they derive values of the ΔH° of $P_4O_6$ as -1695.5 kJ/mol. Smoes and Drowart [24] derive a value for ΔH° of $P_4O_6$ of -1662 kJ/mol using a similar method.

## 2.5 Calculation from vapour pressure

More generally to the sublimation data above, we can calculate the enthalpy of formation of the gas phase if we know the vapour pressure over a solid and the enthalpy of formation of the solid, as follows.

With a solid at equilibrium with its gas phase, there is no net energy in going from solid to gas phase or vice versa.

So ΔG for reaction solid→ gas = 0

$\Delta G_{(g)} = \Delta G_{(s)}$

and hence

$\Delta G°_{(s)} + R \cdot T \cdot \ln(\{solid\}) = \Delta G°_{(g)} + R \cdot T \cdot \ln(\{gas\})$

As the activity of a pure solid is 1, and the activity of the gas approximates to its partial pressure, then this simplifies to

$\Delta G°_{(s)} - \Delta G°_{(g)} = R \cdot T \cdot \ln(P_v)$

Where $P_v$ is the vapour pressure at that temperature. Vapour pressure over pure $P_4O_6$ can be calculated at 298 K from [25] to be $1.359 \times 10^{-9}$ bar. It follows that $\Delta G°_{(g)}$ is 50.586 kJ/mol more positive than the $\Delta G°_{(s)}$ at 298 K. Note this is not a difference in ΔH, as ΔG includes the entropy increase on evaporation.

## 3. Discussion

### 3.1 Free energy of formation of $P_4O_{6(g)}$

We can now compare the $\Delta G°_{(g)}$ for $P_4O_6$ as derived by the different methods above. These results are summarized in Tables 4 through 7.

| Enthalpy of formation in gas phase | Basis of estimate of enthalpy | Section | Inferred $\Delta G°_{(g)}$ (kJ/mol) assuming $\Delta S^f_{(g)} =$ 0·345.61 J/mol/K | Inferred $\Delta G°_{(g)}$ (kJ/mol) assuming $\Delta S^f_{(g)} =$ 0·434.108 J/mol/K |
|---|---|---|---|---|
| -1695.5 | Energy of atomization [23] | 2.4 | -1592.51 | -1566.14 |
| -1662 | Energy of atomization [24] | 2.4 | -1559.01 | -1532.64 |
| -1706.3 | *Ab initio* QM methods [22] | 2.3 | -1603.31 | -1576.94 |

**Table 4.** Gibbs Free energy of formation of $P_4O_6$ estimated from gas phase estimates of enthalpy of formation ($\Delta H°$)

| Enthalpy of formation of solid phase | Basis | Section | Enthalpy of gas phase | $\Delta G°_{(g)}$ (kJ/mol) of gas phase |
|---|---|---|---|---|
| -2270.87 | From solid phase enthalpy from [5], and entropy difference between solid and gas phases | 2.2 | -2163.78 | -2028.55 |
| -1671.1 | From solid phase enthalpy from [8], and entropy difference between solid and gas phases | 2.2 | -1564.01 | -1428.78 |

**Table 5.** Gibbs Free energy of formation of $P_4O_6$ estimated from solid phase enthalpy and entropies of solid and gas phases

| Enthalpy of formation (solid) | Basis | Assuming ΔG°$_{(g)}$ from vapour pressure [25] |
|---|---|---|
| -2270.9 | Solid phase measured directly [15] assuming solid phase entropy = -545.8 | -2057.64 |
| -2270.9 | Solid phase measured directly [15] assuming solid phase entropy = -582.1 | -2046.82 |
| -1671.1 | Solid phase measured directly [8] assuming solid phase entropy = -545.8 | -1457.87 |
| -1671.1 | Solid phase measured directly [8] and assuming solid phase entropy = -582.1 | -1447.05 |

**Table 6.** Gibbs Free energy of formation of $P_4O_6$ estimated and solid phase enthalpy of formation (Section 2.1), entropy (section 2.2), and energy of vapourization (section 2.5)

| Basis | Section | ΔG°$_{(g)}$ (kJ/mol) |
|---|---|---|
| Stated in HSC | 2.1 | -1479.718 |
| Calculated from semi-empirical QM calculations [20] | 2.3 | -1859.79 |
| NIST/JANAF tables [4] | 2.1 | -2084.877 |

**Table 7.** Gibbs Free energy of formation of $P_4O_6$ reported in literature sources

Tables 4 through 7 provide a range of Gibbs standard free energy of formation for $P_4O_6$ in gas phase from -1428.78 to -2084.77 kJ/mol, a range of 656 kJ/mol. We have not presented Gibbs standard free energies calculated from the Soviet era reports of the enthalpy of formation of $P_4O_{6(s)}$ or $P_4O_{6(g)}$ [18] as the original work on which these are based is not accessible; however inclusion of those values would not change our overall conclusions.

## 3.2 Possible sources of the differences in ΔG° values.

The most obvious reason for the differences in ΔG° values for $P_4O_6$ that can be derived from the literature is that the experimental values on which they were based are in error. This is the explanation put forward by [8]. We address this briefly below, but detailed examination of experimental methods is beyond the scope of this summary. Calculations of gas phase values from solid phase data assume that $P_4O_{6(s)}$ is a single substance: however [25] note that there are at least two forms of $P_4O_{6(s)}$ with substantially different vapour pressures; if $P_4O_6$ in the gas phase is always the same, then the heat of formation of these two solid phases must be different. Several authors suggest that gas phase $P_4O_6$ may contain $P_2O_3$ species as well [24], so what 'gas phase' and 'solid phase' $P_4O_6$ mean is poorly defined. We also note that the NIST/JANAF tables are not always accurate, or even internally consistent. For example, we found inconsistencies between PDF and online versions of the NIST/JANAF tables, and between different parameters within those tables, for phosphine [20].

Changes in reference enthalpies may also be a source of difference. For example, Smoes and Drowart's [24] estimate of -1662 kJ/mol becomes -1682.4 kJ/mol when the more up-to-date atomization energies of $P_4$ and $O_2$ from NIST/JANAF [5, 21] are used. Similarly, the solid phase enthalpy from [8] is -1671.06 kJ/mol rather than -1620.51 if the NIST/JANAF value for the solid phase enthalpy of $P_4O_{10}$ is used as reference. But adopting the lower value is only valid if the NIST/JANAF values for $P_4O_{10}$ are accurate.

## 3.2 Observational constraints on the ΔG of $P_4O_6$.

The lowest (most negative) of the $\Delta G°_{(g)}$ values in Tables 4 through 7 can be ruled out observationally. For example, we calculate the equilibrium between P(III) and P(V) under crustal conditions in the Earth's surface, using the free energy values from NIST/JANAF and from HSC, and compare these to three standard rock redox buffers (see [26] for a review of rock redox buffers), and a fourth buffer relevant to phosphorus mineral redox. The results are shown in Figure 1.

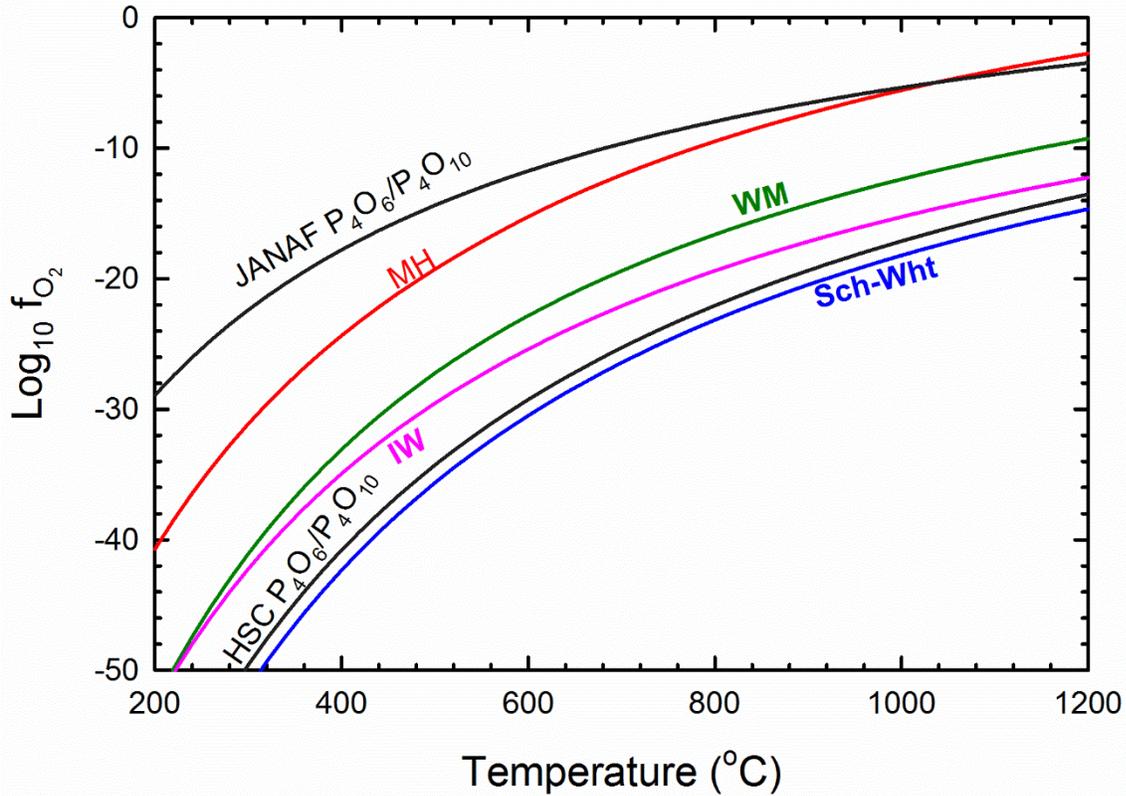

**Figure 1.** Oxygen fugacity of the P(III)↔P(V) pair under assumption of NIST/JANAF thermochemical data (top black line) and HSC data (lower black line), compared to the fugacity of standard redox buffers MH (Magnetite/Haematite), WM (Wustite/Magnetitie) and IW (Iron/Wüstite). Also shown is the buffer associated with oxidation of schreibersite (as $Fe_3P$) to whitlockite (as $Ca_3(PO_4)_2$) via $3CaSiO_3 + 2Fe_3P + 2.5O_2 = Ca_3(PO_4)_2 + 6Fe + 3SiO_2$.

Figure 1 shows that even if the crust is highly oxidized (MH buffer), if we assume the NIST/JANAF values for the thermodynamics of P species are correct then we would expect phosphorus to be present in the crust as P(III), and hence volcanic magma and gases should contain the majority of their phosphorus as $P_4O_6$. $P_4O_6$ would react rapidly with volcanic and atmospheric water to form phosphine. However phosphine is not detected in volcanic gases [27, 28], and one model of the chemistry of phosphorus in gas phase in volcanic erupta suggests it is predominantly P(V) [29]. This implies that the NIST/JANAF value for $\Delta G°_{(g)}$ of $P_4O_6$ must be too low. Similarly, initial laboratories studies reacting $P_4O_{10}$ with a 10-fold excess of $Fe_2O_3/Fe_3O_4$ (haematite: magnetite 1:1) at 500°C confirm that the large majority of the phosphorus remains in P(V) oxidation state under these conditions and is not reduced to P(III) as predicted by the NIST/JANAF data (Feng and Pasek in prep.).

Furthermore, the reaction

$P_4O_{6(s)} + 6H_2O_{(l)} \rightarrow 4H_2PO_3^-{}_{(aq)} + 4H^+{}_{(aq)}$

is a textbook preparative method for phosphorous acid [30] which goes essentially to completion. We assume this means that the equilibrium constant for the reaction is therefore at least 100, so that less than 1% of the starting materials were left in the product, a common-cutoff for a 'clean' preparation of a substance. (For example, Miller [31] found that 'pure' $P_4O_6$ made by oxidizing phosphorus in limiting oxygen contained around 1% unreacted phosphorus.) If the equilibrium constant for the reaction is >100, then the free energy of formation of $P_4O_{6(s)}$ cannot be less (more negative) than -1977 kJ/mol (assuming the thermodynamic values for $H_2O$ and $H_2PO_3^-$ from [32]), and therefore the free energy of formation of $P_4O_{6(g)}$ must be no less (more negative) than -1927 kJ/mol. Other constraints may also exclude the higher values of $\Delta G^0$ for $P_4O_6$ used in HSC's database.

## 3.3 Implications for the formation of PH₃ from P₄O₆ in Venus' atmosphere

It has been suggested that the tentative detection of phosphine (PH$_3$) in Venus' atmosphere [33, 34] can be explained by abiotic processes [3], and specifically by (among other processes) the disproportionation of P$_4$O$_6$, based on the thermodynamic data from the HSC database. Previously [2] had ruled out this reaction as a source of phosphine, and instead P$_4$O$_6$ gas was presumed to be a sink of volatile P in Venus's lower atmosphere based on the thermodynamic values from NIST/JANAF. Here we show that, regardless of the ΔG°$_{(g)}$ for P$_4$O$_6$, the disproportionation of P$_4$O$_6$ cannot explain the presence of ~1 ppb PH$_3$ in Venus' atmosphere.

Bains et al. [2] estimated that P$_4$O$_6$ would be a major species in the lower atmosphere of Venus, but that disproportionation to PH$_3$ was not thermodynamically favoured. Based on the HSC values, Omran et al. [3] estimated that disproportionation of P$_4$O$_6$ to PH$_3$ was favoured; however, the same ΔG values would show that P$_4$O$_6$ could not occur in the atmosphere of Venus, assuming that the atmosphere is in thermodynamic equilibrium with respect to phosphorus species. Specifically, assuming the atmosphere is in thermodynamic equilibrium, the abundance of P$_4$O$_6$ can be calculated from the abundance of H$_2$, H$_3$PO$_4$ and H$_2$O and the free energy of the reaction

H$_3$PO$_4$ + H$_2$ → ¼ P$_4$O$_6$ + 2½ H$_2$O

If the reaction is at equilibrium, then

$$\Delta G = 0 = \Delta G^o + R \cdot T \cdot \ln(\frac{[P_4O_6]^{0.25} \cdot [H_2O]^{2.5}}{[H_3PO_4] \cdot [H_2]})$$

Where [X] is the partial pressure of X (i.e. the mole fraction of the gas in the atmosphere multiplied by the pressure). From this equation we can calculate the abundance of P$_4$O$_6$ assuming abundances of the other species and thermodynamic values. This calculation is summarized in Table 8.

| Substance | Abundance (mole fraction) | Standard free energy of formation (kJ/mol) | | |
|---|---|---|---|---|
| | | NIST/JANAF | HSC | Average |
| $PH_{3(g)}$ | 1.00E-09 | 30.893 | 13.45 | 22.1715 |
| $P_4O_{6(g)}$ | | -2085 | -1480 | -1679 |
| $H_3PO_{4(l)}$ | 3.25E-01 | -1123.6 | -1149 | -1136.3 |
| $H_2O_{(g)}$ | 1.0E-04 | -228.6 | -228.6 | -228.6 |
| $H_{2(g)}$ | 9.00E-10 | 0 | 0 | 0 |
| | | | | |
| Standard free energy of disproportionation reaction | $P_4O_6 + 6H_2O \rightarrow [4\ H_3PO_3] \rightarrow 3\ H_3PO_4 + PH_3$ | 116.69 | -581.95 | -465.26 |
| | | | | |
| Free energy of reduction of $H_3PO_4$ per mole of $P_4O_6$ | $4\ H_3PO_4 + 4H_2 \rightarrow P_4O_6 + 10\ H_2O$ | 123.4 | 746 | 538.2 |
| Inferred abundance of $P_4O_{6(g)}$ at cloud level | | $1.7 \cdot 10^{-20}$ ($1 \cdot 10^{-6} - 6 \cdot 10^{-29}$) | $2.3 \cdot 10^{-144}$ ($1.4 \cdot 10^{-130} - 6 \cdot 10^{-153}$) | $1.4 \cdot 10^{-100}$ ($8.5 \cdot 10^{-87} - 5.1 \cdot 10^{-109}$) |

**Table 8.** Disproportionation of $P_4O_6$ to $PH_3$ is unlikely under any scenario at 55 km on Venus. Top section: column 2 = average abundance of $PH_3$, $H_3PO_4$, $H_2O$ and $H_2$ in Venus atmosphere from [2]. $H_3PO_4$ is assumed to be present as liquid in the cloud droplets, and to comprise 10% of those droplets, following the high phosphorus loading of cloud droplets found by the Vega descent probe X-ray fluorescence analyzer [35]. Columns 3 thru 5 = standard free energy of formation of substances from NIST/JANAF, HSC, and an average of the two. Central section: Standard free energy of the disproportionation reaction derived from these values, showing that the reaction is disfavoured under NIST/JANAF assumptions, favoured under other assumptions. Bottom panel: calculation of abundance of $P_4O_6$ assuming the abundances of $H_3PO_4$, $H_2$ and $H_2O$ and the free energy of formation of all compounds; top row = free energy of reaction reducing $H_3PO_4$ to $P_4O_6$ given the $\Delta G$ terms from different sources, bottom row = atmospheric abundance inferred from this reaction $\Delta G$, with ranges for water values $4.2 \times 10^{-6}$ to $7 \times 10^{-4}$

Table 8 summarizes our calculations based on NIST/JANAF values [4, 5], on the HSC values [6], and an average of the two as an illustrative third 'compromise' case. Bains et al. [2] predicted there would be vanishingly small amounts of $P_4O_6$ in the atmosphere at 55 km using the NIST/JANAF $\Delta G°_{(g)}$ values (they predicted $9.7 \cdot 10^{-19}$ of the phosphorus in gas phase at 55 km would be present as $P_4O_6$), but that could be enough to allow $P_4O_6$ to be a trace intermediate in the atmospheric chemistry, and hence be a substrate for disproportionation reactions. However, Bains et al predicted that the disproportionation reaction is not thermodynamically favored. Under Omran et al's [3] assumptions disproportionation is thermodynamically favored but there is no $P_4O_6$ present to disproportionate. (Note that these thermodynamic values are calculated based on an assumed abundance of $PH_3$ of 1 ppb, which is what Bains et al. and Omran et al. were seeking to explain.)

Figure 2 generalizes this calculation to all the values of $\Delta G°_{(g)}$ for $P_4O_6$ listed in Tables 4 through 7, for water abundance values of $4.2 \times 10^{-6}$, $1 \times 10^{-4}$ and $7 \times 10^{-4}$ [36], and for temperatures and pressures in the middle of the clouds and the lower haze layer below the clouds.

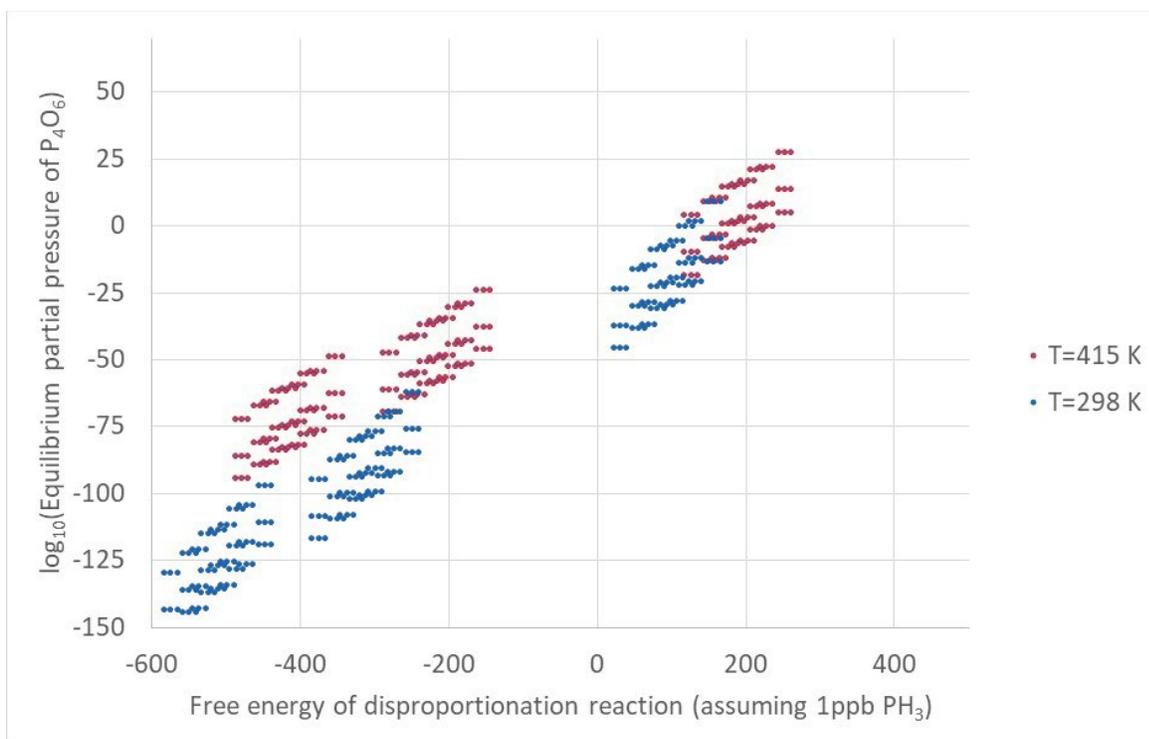

**Figure 2.** There is no combination of thermodynamic values that allows $PH_3$ to be made by disproportionation of $P_4O_6$ at cloud levels. X axis: Free energy of the reaction $P_4O_6 + 6\ H_2O \rightarrow PH_3 + 3\ H_3PO_4$ (kJ/mol) assuming 1 ppb $PH_3$. Y axis: Equilibrium concentration of $P_4O_6$ in gas phase. Plotted are the values calculated from all possible combinations of the thermodynamic values listed in Table 2, and for water abundances of $4.2\times10^{-6}$, $1\times10^{-4}$ and $7\times10^{-4}$. Results are plotted for values at 298 K and at 415 K, the latter assuming the entropy of formation values from the NIST/JANAF data [5].

Figure 2 shows that there is no combination of thermodynamic values that allows $P_4O_6$ to be present *and* allows $PH_3$ to be formed by disproportionation of $P_4O_6$ and to accumulate to 1 ppb, at 298 K (the temperature of the mid-cloud layer) or at 415 K (the temperature at 40 km altitude, in middle of the haze layer below the clouds). This rules out $P_4O_6$ disproportionation as a source of $PH_3$ in Venus clouds, although it does not rule out other abiotic sources for $PH_3$ [3].

## 3.4 Future work

Given the importance of phosphorus chemistry to the study of atmosphere of solar system gas giants, of Venus, and of exoplanets [37], the accurate thermodynamics of $P_4O_6$ (and other phosphorus species) in the gas phase needs to be resolved. The importance of resolution lies not only in equilibrium calculations for atmospheres, but also for kinetic models. Missing values in kinetic networks are commonly estimated from equilibrium thermodynamics; if thermodynamic parameters are not accurate, then the derived kinetic constants will also be inaccurate, and can result in substantial differences in the predictions of kinetic models (e.g. see [37]). Phosphorus chemistry is not the only such case, and the need for accurate measurements in astronomy, geology and astrobiology has been argued before [38]. Although our interest in the thermodynamics of $P_4O_6$ comes from an interest in the phosphorus chemistry of Venus' atmosphere, the discrepancies also have important impact on the studies of gas giant, hot Jupiter and brown dwarf atmospheres [7, 9-11, 37]. Although we have not identified any geological chemistry papers that cite [8], we expect that calculations of phosphorus speciation in the mantle (e.g. [27]) will also be substantially affected by differences in the $\Delta G°$ assumed for $P_4O_6$.

We advise that the thermodynamic values for all the phosphorus species of importance to atmospheric and geochemistry be recalculated from original data using reference data derived from the most accurate sources, with the accuracy of the original measurements evaluated by a group that is personally familiar with the methods involved. (This is not to denigrate others who use these values for modelling, but experimental methods often have subtleties that only practitioners appreciate. To take just one example, Hartley and McCoubrey [8] derive the enthalpy of formation for $P_4O_6$ from the enthalpy of complete combustion of $P_4O_6$ to $P_4O_{10}$. They state that their sample of $P_4O_6$ was made by combustion of white phosphorus in limiting oxygen as described by Thorpe and Tutton [39]; however Miller [31] points out that Thorpe and Tutton's method results in a material with ~1% elemental phosphorus in solid solution, even after extensive purification. If the material used by Hartley and McCoubrey contained trace elemental phosphorus, the apparent enthalpy of combustion would be larger than for pure $P_4O_6$.) In particular, it would be worthwhile checking that the conventions used in calculation are consistent. We note that the NIST/JANAF values, for example, are calculated according to the Berman-Brown convention [40], wherein $\Delta G°$ is calculated relative to the $G°$ of elements at 298K and 1 bar, and so does not account for phase and specific heat capacity changes in reference elements with temperature. While internally consistent, this convention can result in significantly different $\Delta G°$ values to those calculated from more complete treatments, and so mixing the conventions can result in substantial errors in free energies calculated at high temperatures (such as volcanic mineralogy or the deep atmospheres of Jovian planets: e.g. see [41]). However the differences in convention should not affect the $\Delta G°$ values calculated in this paper, which by definition are calculated for 298 K, and should have minimal effect on conclusions about phosphorus chemistry in Venus' mid-cloud altitudes, where temperatures range from ~270 - ~320 K [42].

New accurate measurements using modern analytical techniques of the enthalpy of the reactions

$P_4 + 3O_2 \rightarrow P_4O_6$

$P_4 + 5O_2 \rightarrow P_4O_{10}$

$P_4O_6 + 2O_2 \rightarrow P_4O_{10}$

would be a valuable addition but are not trivial measurements to make. An ideal would be to measure the equilibrium position in any of the reactions above as a function of temperature, possibly using standard geological oxygen fugacity buffers as mentioned above (Feng and Pasek in prep.). Absent accurate measurement, detailed ab initio QM calculations may help to resolve the uncertainties.

Finally, we note that we have been specifically motivated to reveal uncertainties in the thermodynamics phosphorus gas phase chemistry. We encourage those with interests in other elements that have complex atmospheric chemistry, such as sulfur and the halogens, to explore whether similar uncertainties in the thermodynamics of these species exist buried in the assumptions behind current models.

## Acknowledgements

We are grateful to Anna Markovska for help with Russian translation. We are grateful to our reviewer, who made a number of constructive comments that improved this paper. MAP was supported by a grant from NASA Emerging Worlds (80NSSC23K0019).